\documentclass[a4paper, oneside, twocolumn, notitlepage, 10pt]{extarticle_ecoc}
\usepackage{ecoc}
\usepackage{graphicx}
\usepackage{amsmath}
\usepackage{tikz}
\usepackage{pgfplots}
\usepackage{mathtools}
\usepgfplotslibrary{colorbrewer}
\usepgfplotslibrary{groupplots}
\usetikzlibrary{shapes.geometric}
\usepackage{pgfplotstable}
\pgfplotsset{compat=1.17}
\usetikzlibrary{backgrounds}
\usepackage{wrapfig}  
\usepackage{subfigure}
\usepackage{amsfonts}
\usepackage{multirow}
\usetikzlibrary{arrows}
\usetikzlibrary{decorations.markings}
\usetikzlibrary{matrix}
\usepackage{xcolor}
\usepackage{colortbl}
\usepackage{threeparttable}

\newcommand{\ts}{\textsuperscript}

\begin{document}
\selectlanguage{english}    


\title{A Modulation-Format Dependent Closed-form Expression for the Gaussian Noise Model in the Presence of Raman Amplification}%


\author{
    H.~Buglia\textsuperscript{(1)}, M.~Jarmolovi\v{c}ius\textsuperscript{(1)},
    L.~Galdino\textsuperscript{(2)}, R.I. Killey\textsuperscript{(1)}, and~P.~Bayvel\textsuperscript{(1)}
}

\maketitle                  


\begin{strip}
 \begin{author_descr}

   \textsuperscript{(1)} Optical Networks Group, Department of Electronic and Electrical Engineering, UCL (University College London), Torrington Place, London, WC1E 7JE, United Kingdom,
   \textcolor{blue}{\uline{henrique.buglia.20@ucl.ac.uk}}.

   \textsuperscript{(2)} Corning Optical Communications, Ewloe, CH5 3XD, United Kingdom.

 \end{author_descr}
\end{strip}

\setstretch{1.1}
\renewcommand\footnotemark{}
\renewcommand\footnoterule{}


\begin{strip}
  \begin{ecoc_abstract}

A closed-form expression that estimates the nonlinear interference of arbitrary modulation formats in Raman amplified links is presented. Accounting for any pumping schemes and inter-channel stimulated Raman scattering effect, the formula is applied to an optical bandwidth of 20~THz and validated using numerical simulations. \textcopyright2023 The Author(s)


  \end{ecoc_abstract}
\end{strip}


\section{Introduction}

Raman amplification (RA) has attracted considerable attention in recent years because of its potential to achieve increased data rates in optical fibre transmission systems~\cite{Souza:22}. For ultra-wideband (UWB) systems, this technique can be used in combination with lumped amplification (e.g., rare-earth-doped fibre or semiconductor optical amplifiers) to unlock additional transmission capacity by compensating for the worst noise figure of these amplifiers~\cite{Buglia:22}. This combination, also called hybrid amplification, has been used over the past few years to achieve milestones of data throughput in single-mode fibres over different distances~\cite{Puttnam:22,longhaul}.

For systems operating fully or in combination with Raman amplification, the estimation of the system performance can be carried out using nonlinear models, such as the widely used Gaussian Noise~(GN) model in integral form~\cite{gnmodel,isrsgnmodel,Carena:14}. However, the complexity of this estimation rapidly scales with the transmission bandwidth and closed-form formulas of this model must be used to achieve a real-time prediction of the system performance. 

To date, the only closed-form expressions of the GN model accounting for forward and backward Raman amplification, inter-channel stimulated Raman scattering (ISRS) and validated over UWB systems is the one published in~\cite{bugliaOFC,arxiv}. This closed-form expression is valid for Gaussian-modulated signals but does not account for arbitrary modulation formats such as quadrature amplitude modulation (QAM). 

In this work, a modulation-format correction term is derived, enabling the formula in~\cite{bugliaOFC,arxiv} to account for any modulated signal. The new closed-form expression is then validated for signal transmission over a 20~THz optical bandwidth, making full use of S-, C- and L- bands. The simulation scenario is for a 64-QAM signal in a hybrid amplification scheme, where the S-band is fully amplified by launching backward Raman pumps in the transmission fibre. The accuracy of the formula is validated through split-step Fourier method (SSFM) simulations using a large $2^{18}$ symbol sequence per channel.
\begin{figure}[t!]
 \centering
  \begin{tikzpicture}[baseline]
\begin{axis}[yshift=-0.2cm,
legend cell align=left,
title={\footnotesize (a)},
title style={at={(axis cs:1485,0.282)}},
legend style={font=\footnotesize, at={(rel axis cs:0,1)}, anchor=north west},
width=8cm, height = 4.5cm,
xlabel={Wavelength [nm]},
ylabel={Attenuation [dB/km]},
grid=both,
ytick={0.19,0.21,0.23,0.25,0.27,0.29},
xtick={1350,1390,1430,1470,1510,1550,1590,1630},
ymax=0.29,ymin=0.19,
xmin=1350,xmax=1630,
xticklabel style={/pgf/number format/1000 sep=},
    ytick distance=0.02,
    xtick distance=40,
]

\addplot[black,very thick] table[x=wavelength,y=loss] {Data/ECOC_attenuation_profile.txt};

\node[anchor=west] (source) at (axis cs:1428-25,0.28){E};
\node[anchor=west] (source) at (axis cs:1493-10,0.28){S};
\node[anchor=west] (source) at (axis cs:1543-6,0.28){C};
\node[anchor=west] (source) at (axis cs:1588-2,0.28){L};

\begin{scope}[on background layer]
    \fill[red,opacity=0.1] ({rel axis cs:0.035714,0.0}) rectangle ({rel axis cs:0.392857,1});
    \fill[green,opacity=0.1] ({rel axis cs:0.392857,0}) 
    rectangle ({rel axis cs:0.642857,1});
    \fill[blue,opacity=0.1] ({rel axis cs:0.642857,0})
    rectangle ({rel axis cs:0.767857,1});
    \fill[gray,opacity=0.1] ({rel axis cs:0.767857,0})
    rectangle ({rel axis cs:0.98214285,1});
\end{scope}

\end{axis}
\end{tikzpicture}
\begin{tikzpicture}[baseline]
\begin{axis}[yshift=-0.2cm,
title={\footnotesize (b)},
title style={at={(axis cs:13.5,0.46)}},
legend cell align=left,
legend style={font=\footnotesize, at={(rel axis cs:0,1)}, anchor=north west},
width=8cm, height = 4.5cm,
xlabel={Frequency separation [THz]},
ylabel={Raman gain [1/km/W]},
grid=both,
ymax=0.5,ymin=0,
xmin=0,xmax=26,
    ytick distance=0.1,
    xtick distance=4,
]

\addplot[Set1-C,thick] table[x=Delta_Freq,y=Raman_Gain] {Data/ECOC_Raman.txt};

\end{axis}
\end{tikzpicture}
\caption{ (a) Attenuation coefficient and (b) Raman gain spectrum of an ITU-T G652.D fibre.}\label{fig:raman_loss}
\end{figure}
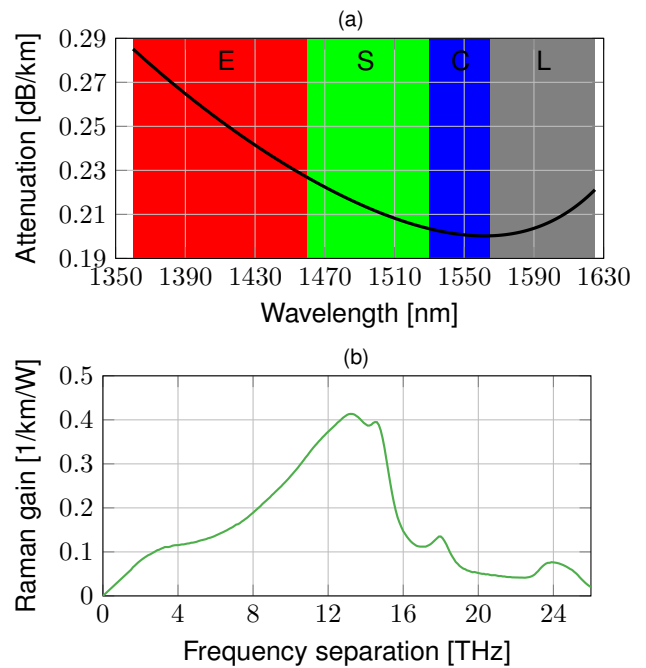

\section{The closed-form expression}
\begin{figure*}[h]
\vspace*{0.5cm}
\begin{equation}\tag{2}\label{eq:NLI}\end{equation}
\vspace*{-2.3cm}
\begin{equation*}
	\resizebox{\textwidth}{!}{$%
	\begin{split}
 &\eta_{\text{corr,n}}(f_i) =  \frac{80}{81} \sum_{k = i, k \neq i}^{N_{ch}} \Phi_k \frac{\gamma^2}{B_k} \left(\frac{P_k}{P_i}\right)^2 \sum_{\substack{0 \leq l_1 + l_2 \leq 1 \\ 0 \leq l_1^\prime + l_2^\prime \leq 1}}\Upsilon_k \Upsilon_k^\prime \left\{\frac{2(\kappa_{f,k} \kappa_{f,k}^\prime + \kappa_{b,k} \kappa_{b,k}^\prime)}{\phi_{i,k}(\alpha_{l,k} + \alpha_{l,k}^\prime)} \left[\mathrm{atan} \left(\frac{\phi_{i,k}B_i}{2\alpha_{l,k}}\right)	+\mathrm{atan}\left(\frac{\phi_{i,k}B_i}{2\alpha_{l,k}^\prime}\right)\right] \right.+\\
&+ \pi \left[ - \frac{\kappa_{f,k} \kappa_{b,k}^\prime +  \kappa_{b,k} \kappa_{f,k}^\prime}{\phi_{i,k}(\alpha_{l,k} + \alpha_{l,k}^\prime)} \left(  \mathrm{sign}\left(\frac{\alpha_{l,k}}{\phi_{i,k}} \right)  e^{-|\alpha_{l,k}L|} + 
 \mathrm{sign}\left(\frac{\alpha_{l,k}^\prime}{\phi_{i,k}} \right)  e^{-|\alpha_{l,k}^\prime L|}\right) + \frac{\kappa_{f,k} \kappa_{b,k}^\prime - \kappa_{b,k} \kappa_{f,k}^\prime}{\phi_{i,k}(\alpha_{l,k} + \alpha_{l,k}^\prime)} \times \right. 
 \\
&\times \left. \left.
	\left(  \mathrm{sign}\left(-\phi_{i,k} \right)  e^{-|\alpha_{l,k}L|} +  \mathrm{sign}\left(\phi_{i,k} \right)  e^{-|\alpha_{l,k}^\prime L|}\right) \vphantom{\frac{\kappa_{f,k}^\prime}{\kappa_{f,k}^\prime}} \right]+ \left[ \frac{2\pi \tilde{n} (\kappa_{f,k} - \kappa_{b,k})(\kappa_{f,k}^\prime - \kappa_{b,k}^\prime)}{|\tilde{\phi_k}|B_k^2 \alpha_{l,k} \alpha_{l,k}^\prime} 
    \cdot\left((2\Delta f - B_k) \ln{\left(\frac{2\Delta f - B_k}{2\Delta f + B_k}\right)} + 2B_k \right) \right ] \right\}
 	\end{split}$%
	}
\end{equation*}
\vspace*{-0.5cm}

\end{figure*}

The signal-to-noise ratio ($\text{SNR}_{i}$) for the $i$-th channel of interest (COI) at the end of the $n$-th span after amplification can be estimated as $\text{SNR}_{i}^{-1} \approx \text{SNR}_{\text{NLI},i}^{-1} + \text{SNR}_{\text{ASE},i}^{-1}+ \text{SNR}_{\text{TRX},i}^{-1}$, where 
$\text{SNR}_{\text{NLI},i}$, $\text{SNR}_{\text{ASE},i}$ $\text{SNR}_{\text{TRX},i}$ originate from fibre nonlinear interference (NLI), amplifier noise and transceiver noise, respectively. This work is devoted to the calculation of $\text{SNR}_{\text{NLI},i} \approx \frac{P_i}{\eta_{\text{n}}(f_i)P_i^3}$,
where $P_i$ is the launch power of the COI and $\eta_{\text{n}}(f_i)$ is its nonlinear coefficient, which is further decomposed as $\eta_{\text{n}}(f_i) = \eta_{\text{GN,n}}(f_i) + \eta_{\text{corr,n}}(f_i)$, where $\eta_{\text{GN,n}}(f_i)$ and $\eta_{\text{corr,n}}(f_i)$ are respectively the nonlinear coefficient contributions accounting for Gaussian modulated symbols~\cite{isrsgnmodel,arxiv} and its correction term accounting for the dependence of the NLI on the modulation format~\cite{Carena:14,ISRSGNmodel_correction}. $\eta_{\text{GN,n}}(f_i)$ is calculated in closed-form using Eqs.~(11) and (12) in~\cite{arxiv}. For Gaussian modulated
signals, the correction term $\eta_{corr,n}(f_i)$ vanishes and one obtains the model in~\cite{isrsgnmodel,arxiv}. In this work, we complement the work in~\cite{arxiv} by deriving a closed-form expression for $\eta_{corr,n}(f_i)$. 

The modulation format correction term contribution of the nonlinear coefficient is the sum of all COI-interfering-channel pairs present in the transmitted signal, given in integral form by Eqs.~(9) and (12) in~\cite{ISRSGNmodel_correction}, i.e,
\vspace*{0.7cm}
\begin{equation}\label{eq:eta_corr}\end{equation}
\vspace*{-1.7cm}
\begin{equation*}
\resizebox{.45\textwidth}{!}{$%
\begin{aligned}
    &\eta_{corr,n}(f_i) = \frac{80}{81} \sum_{k=1,k \neq i}^{N_{\text{ch}}} \left(\frac{P_k}{P_i}\right)^2 \frac{\gamma^2 \Phi_k}{B_k} \times \\ 
    &\times \left\{ \int_{\frac{-B_i}{2}}^{\frac{B_i}{2}} |\mu (f_1 + f_i,f_k,f_i)|^2 df_1 \
    + |\mu (f_i,f_k,f_i)|^2 \times \right.\\ 
    & \left. \times \frac{2\pi \tilde{n}}{| \tilde{\phi_k}|B_k^2} (2\Delta f - B_k) \ln{\left(\frac{2\Delta f - B_k}{2\Delta f + B_k}\right)} + 2B_k \right\}
\end{aligned}$}%
\end{equation*}
where $\tilde{n} = 0$ for a single span or $\tilde{n} = n$ otherwise. $B_k$ is the bandwidth of the channel $k$, $\mu$ is the so-called link function which is given in closed-form as Eq.~(9) in~\cite{arxiv}. $\gamma$ is the nonlinear parameter, $\Delta f = f_k - f_i$ is the frequency separation between channel $i$ and $k$ with frequencies $f_i$ and $f_k$, $\tilde{\phi_k}=-4\pi^2\left[\beta_2+\pi\beta_3(f_i+f_k\right)]L$, where $\beta_2$ is the group velocity dispersion parameter and $\beta_3$ its
linear slope and $L$ is the span length. 
$\Phi_k$ stands for the excess kurtosis of the given constellation, providing statistical characteristics of the signal, and reflecting how the constellation deviates from the Gaussian one. 

Let $T_{f,k} = -\frac{P_fC_{f,k}(f_k-\hat{f})}{\alpha_{f,k}}$, $T_{b,k} = -\frac{P_bC_{b,k}(f_k-\hat{f})}{\alpha_{b,k}}$, $T_k = 1 + T_{f,k} - T_{b,k}e^{-\alpha_{b,k}L}$, $\alpha_{l,k} = \alpha_k + l_1 \alpha_{f,k} - l_2 \alpha_{b,k}$, $\kappa_{f,k} = e^{-(\alpha_k + l_1\alpha_{f,k})L}$, $\kappa_{b,k} = e^{-l_2\alpha_{b,k} L}$, $\Upsilon_k
= T_k \left(\frac{-\tilde{T}_{f,k} }{T_k} \right)^{l_1} \left(\frac{\tilde{T}_{b,k} }{T_k} \right)^{l_2}$.\\ A closed-form expression of Eq~\eqref{eq:eta_corr} is given by Eq.~\eqref{eq:NLI}, where $\phi_i=-4\pi^2\left(\beta_2+2\pi\beta_3f_i\right)$, $\phi_{i,k}=-4\pi^2\left(f_k-f_i\right)\left[\beta_2+\pi\beta_3\left(f_i+f_k\right)\right]$ and $\hat{f}$ is the average frequency of the pumps. The coefficient $\Upsilon_k^\prime$ is the same as $\Upsilon_k$ with the indices $l_1$ and $l_2$ replaced by $l_1^\prime$ and $l_2^\prime$. The same is valid for the variables $\alpha_{l,k}^\prime$, $\kappa_{f,k}^\prime$ and $\kappa_{b,k}^\prime$.

The coefficients $\alpha_k$, $C_{f,k}$, $C_{b,k}$, $\alpha_{f,k}$, and $\alpha_{b,k}$ are channel-dependent parameters and matched using nonlinear least-squares fitting to correctly reproduce the solution of the Raman differential equations in the presence of RA (see Eq.~(1) in~\cite{arxiv}) using the profiles in Fig.~\ref{fig:raman_loss}, obtained as a semi-analytical solution of these equations, given by Eq.~(2) in~\cite{arxiv}. Finally, to obtain Eq~\eqref{eq:NLI}, the link-function $\mu$ in closed-form is used as Eq.~(9) in~\cite{arxiv} and the integral in Eq~\eqref{eq:eta_corr} is solved similarly to that in Appendix C of~\cite{arxiv} - the proof is omitted here.


\begin{table*}[t]
\vspace{-0.5cm}
\caption{Pumps' power and wavelength allocation which yields the power profile shown in Fig~\ref{fig:profile_3D}.}
\centering
\label{tab:pumps}
\begin{tabular}{|c|c|c|c|c|c|c|c|c|}
\hline
\multicolumn{1}{|c|}{Wavelength {[}nm{]}}      & \multicolumn{1}{c|}{\cellcolor[HTML]{FFD0D0}1360} & \multicolumn{1}{c|}{\cellcolor[HTML]{FFD0D0}1365} & \multicolumn{1}{c|}{\cellcolor[HTML]{FFD0D0}1370} & \multicolumn{1}{c|}{\cellcolor[HTML]{FFD0D0}1375} & \multicolumn{1}{c|}{\cellcolor[HTML]{FFD0D0}1380} & \multicolumn{1}{c|}{\cellcolor[HTML]{FFD0D0}1385} & \multicolumn{1}{c|}{\cellcolor[HTML]{FFD0D0}1390} & \multicolumn{1}{c|}{\cellcolor[HTML]{FFD0D0}1415} \\ \hline
\multicolumn{1}{|c|}{Pump power {[}mW{]}} & \multicolumn{1}{c|}{50}                           & \multicolumn{1}{c|}{250}                          & \multicolumn{1}{c|}{250}                          & \multicolumn{1}{c|}{250}                          & \multicolumn{1}{c|}{249}                          & \multicolumn{1}{c|}{76}                           & \multicolumn{1}{c|}{158}                          & \multicolumn{1}{c|}{250}                          \\ \hline
\end{tabular}

\begin{tablenotes}\scriptsize
\item[1] \ts{1} The pumps' power at the beginning of the fibre $(z = 0)$ were found to be 0.0016, 0.0155, 0.0279, 0.0490, 0.0817, 0.0394, 0.1188 and 1.9469~mW, respectively from the lowest (1360~nm) to the highest (1415~nm) wavelength.
\end{tablenotes}
\end{table*}

\section{Transmission System and Results}

\begin{figure}[b!]
  \begin{tikzpicture}[baseline]
    \begin{axis}[
    unbounded coords=jump,view={65}{20}, grid=both,
    legend cell align=left,
legend style={font=\footnotesize, at={(rel axis cs:0.7,1,1.0)}, anchor=north},
width=7.5cm,
    x label style={rotate=-35},
     y label style={rotate=8},
    xlabel={Distance [km]},
ylabel={Wavelength [nm]},
zlabel={Power profile [dBm]},
yticklabels={},
x tick label style = {text width = 1.0cm, align = center, rotate = 70},
extra x ticks={25,75},
extra y ticks={1450,1510,1570,1630},
zmin=-20,zmax=5,
ymin=1450,ymax=1630,
xmin=0,xmax=100,
xticklabel style={/pgf/number format/1000 sep=},
yticklabel style={/pgf/number format/1000 sep=},
  ytick distance=200,
  ztick distance=5,
    ]
    
\node[anchor=north east] at (10,1490,5.5) {S};
\node[anchor=north east] at (10,1550,5.5) {C};
\node[anchor=north east] at (10,1600,5.5) {L};
      \addplot3[surf,color=blue,opacity=0.6] file {Data/ECOC_power_3D.txt};
    \end{axis}
\end{tikzpicture}
\caption{Per-channel launch power evolution along the fibre span.}
\label{fig:profile_3D}
\end{figure}
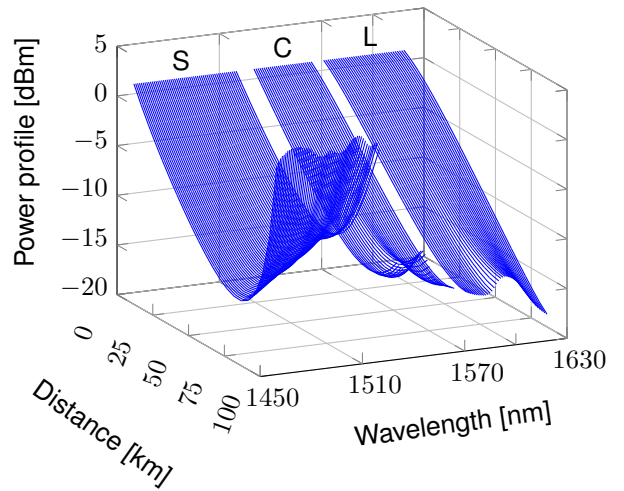

The transmission system under investigation consists of $N_{\text{ch}}$=135 WDM channels with the symbol rate of 140~GBd, spaced by 150~GHz and centred at 1530~nm. A total bandwidth of 20~THz~(160~nm) with spectral gaps of 10~nm and 5~nm between the S/C and C/L bands, respectively, was generated. Each channel was modulated using 64-QAM and Gaussian constellations.
The transmission link comprises 1 or 10 spans of 100km  ITU-T G652.D fibre with attenuation and Raman gain profiles shown in Fig.~\ref{fig:raman_loss}. A spectrally uniform
launch power profile, with 1~dBm power per channel is considered. 
The optical fibre nonlinear coefficient and dispersion parameters are $\gamma = 1.2\text{ W}^{-1} \text{km}^{-1}$,  $D = 16.5~\text{ps }\text{nm}^{-1}\text{km}^{-1}$, and $S = 0.09~\text{ps }\text{nm}^{-2}\text{km}^{-1}$, respectively. 

The link is amplified using hybrid amplification, consisting of a fully backward distributed RA in the S-band to completely recover the transmitted signal, followed by a gain-flattening filter to shape the amplified signal to the input signal. As the RA is not optimised to fully amplify all the bands, ideal lumped amplification is also used to amplify the remaining signals in the C- and L- bands. An example of a full RA system can be found in~\cite{bugliaOFC}.

For RA, eight backward pumps are placed in the E-band, and their wavelengths are chosen to give the greatest gain in the S-band (see  Table~\ref{tab:pumps}). The pump powers were limited to a maximum of 250~mW and they were optimised such that the S-band signal is completely recovered at the receiver - this is set as a nonlinear constraint in the optimisation algorithm. The cost function considered is $\sum_{p} P_p$, such that the power of each pump ($P_p$) is minimised. The pumps' wavelengths and powers are shown in Table~\ref{tab:pumps} and the power profile evolution along the fibre span is shown in Fig~\ref{fig:profile_3D}. The latter is obtained by solving the Raman differential equations given by Eq.~(1) in~\cite{arxiv}, with the pumps shown in Table~\ref{tab:pumps}. Note that, the highest-wavelength pump is 45~nm (6.5~THz) away from the lowest-wavelength channel, such that nonlinearity-induced products generated by the pumps falling within the signal band could be neglected~\cite{Iqbal:20}. 


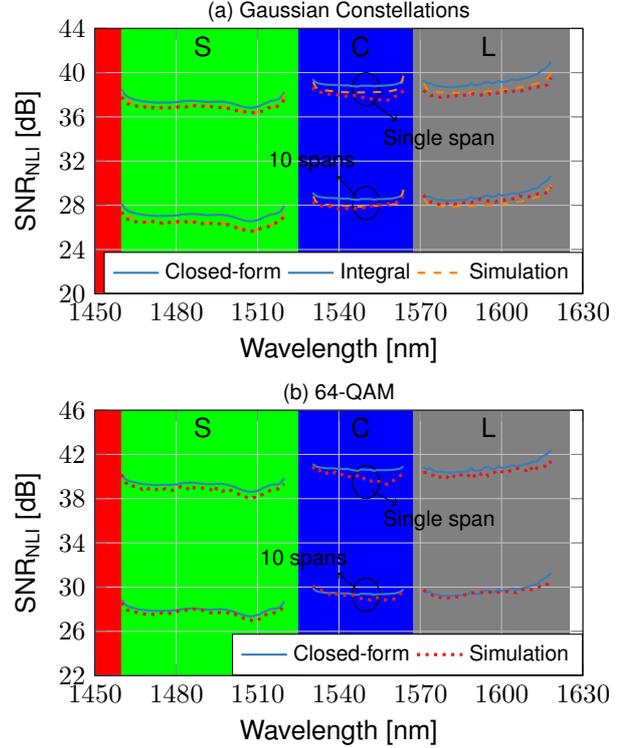
\begin{figure}[t]
\vspace*{-.6cm}
\pgfplotsset{simulation/.style={very thick}}
\pgfplotsset{integral/.style={thick}}
\pgfplotsset{closedform/.style={thick}}

\begin{tikzpicture}[baseline]
\begin{axis}[
clip=false,
unbounded coords=jump,
title = \footnotesize (a) Gaussian Constellations,
title style={at={(axis cs:1540,42.3)}},
legend style={font=\footnotesize, at={(rel axis cs:1,0)}, 
legend columns=3,anchor=south east},
width=8cm, height = 5.1cm,
xlabel={Wavelength [nm]},
ylabel={$\text{SNR}_{\text{NLI}}$ [dB]},
grid=both,
xtick={1450,1480,1510,1540,1570,1600,1630},
ymax=44,ymin=20,
xmin=1450,xmax=1630,
xticklabel style={/pgf/number format/1000 sep=},
  ylabel near ticks,
    ytick distance=4,
    xtick distance=30,
]

\node[anchor=west] (source) at (axis cs:1483,42.4){S};
\node[anchor=west] (source) at (axis cs:1541,42.4){C};
\node[anchor=west] (source) at (axis cs:1589,42.4){L};

           \draw [black] \pgfextra{
\pgfpathellipse{\pgfplotspointaxisxy{1550}{38.5}}
{\pgfplotspointaxisdirectionxy{5}{0}}
{\pgfplotspointaxisdirectionxy{0}{1.5}}
}; 

\node[anchor=west] (source) at (axis cs:1545,38){};
\node (destination) at (axis cs:1565,35){};
\draw[->](source)--(destination);

\node[] at (axis cs: 1577,34) {\footnotesize Single span}; 

           \draw [black] \pgfextra{
\pgfpathellipse{\pgfplotspointaxisxy{1550}{28.2}}
{\pgfplotspointaxisdirectionxy{5}{0}}
{\pgfplotspointaxisdirectionxy{0}{1.5}}
}; 
\node[] at (axis cs: 1530,32) {\footnotesize 10 spans}; 

\node[anchor=west] (source) at (axis cs:1546,28){};
\node (destination) at (axis cs:1536,31.5){};
\draw[->](source)--(destination);

\begin{scope}[on background layer]
    \fill[red,opacity=0.1] ({rel axis cs:0.0,0.0}) rectangle ({rel axis cs:0.055555555,1});
    \fill[green,opacity=0.1] ({rel axis cs:0.0555555,0.0}) rectangle ({rel axis cs:0.41666666,1});
    \fill[blue,opacity=0.1] ({rel axis cs:0.41666666,0}) 
    rectangle ({rel axis cs:0.65277777,1});
    \fill[gray,opacity=0.1] ({rel axis cs:0.652777777,0})
    rectangle ({rel axis cs:0.97222222,1});
\end{scope}

\addlegendentry{Closed-form}
\addplot[Set1-B,closedform] table[x=wavelength,y=closed] {Data/ECOC_BW_SNR_NLI_Gaussxx10.txt};
\addplot[Set1-B,closedform] table[x=wavelength,y=closed] {Data/ECOC_BW_SNR_NLI_Gaussxx1.txt};

\addlegendentry{Integral}
\addplot[Set1-E,integral,dashed] table[x=wavelength,y=integral] {Data/ECOC_BW_SNR_NLI_Gauss.txt};
\addplot[Set1-E,integral, dashed,forget plot] table[x=wavelength,y=integral] {Data/ECOC_BW_SNR_NLI_Gauss_1span.txt};

\addlegendentry{Simulation}
\addplot[Set1-A,simulation,dotted] table[x=wavelength,y=gauss_span1] {Data/ECOC_SSFM.txt};
\addplot[Set1-A,simulation,dotted] table[x=wavelength,y=gauss_span10] {Data/ECOC_SSFM.txt};
%

\end{axis}
\end{tikzpicture}

\begin{tikzpicture}[baseline]
\begin{axis}[
unbounded coords=jump,
title = \footnotesize (b) 64-QAM,
title style={at={(axis cs:1540,44.3)}},
legend style={font=\footnotesize, at={(rel axis cs:1,0)}, legend columns=2,anchor=south east},
width=8cm, height = 5.1cm,
xlabel={Wavelength [nm]},
ylabel={$\text{SNR}_{\text{NLI}}$ [dB]},
grid=both,
xtick={1450,1480,1510,1540,1570,1600,1630},
ytick={22,26,30,34,38,42,46},
ymax=46,ymin=22,
xmin=1450,xmax=1630,
xticklabel style={/pgf/number format/1000 sep=},
/pgf/number format/fixed,
  ylabel near ticks,
    ytick distance=4,
    xtick distance=30,
]
\begin{scope}[on background layer]
    \fill[red,opacity=0.1] ({rel axis cs:0.0,0.0}) rectangle ({rel axis cs:0.055555555,1});
    \fill[green,opacity=0.1] ({rel axis cs:0.0555555,0.0}) rectangle ({rel axis cs:0.41666666,1});
    \fill[blue,opacity=0.1] ({rel axis cs:0.41666666,0}) 
    rectangle ({rel axis cs:0.65277777,1});
    \fill[gray,opacity=0.1] ({rel axis cs:0.652777777,0})
    rectangle ({rel axis cs:0.97222222,1});
\end{scope}
    
\node[anchor=west] (source) at (axis cs:1483,44.4){S};
\node[anchor=west] (source) at (axis cs:1541,44.4){C};
\node[anchor=west] (source) at (axis cs:1589,44.4){L};

           \draw [black] \pgfextra{
\pgfpathellipse{\pgfplotspointaxisxy{1550}{39.5}}
{\pgfplotspointaxisdirectionxy{5}{0}}
{\pgfplotspointaxisdirectionxy{0}{1.5}}
}; 

\node[anchor=west] (source) at (axis cs:1545,39.2){};
\node (destination) at (axis cs:1565,37){};
\draw[->](source)--(destination);

\node[] at (axis cs: 1577,36) {\footnotesize Single span}; 

           \draw [black] \pgfextra{
\pgfpathellipse{\pgfplotspointaxisxy{1550}{29.3}}
{\pgfplotspointaxisdirectionxy{5}{0}}
{\pgfplotspointaxisdirectionxy{0}{1.5}}
}; 
\node[] at (axis cs: 1527,32.5) {\footnotesize 10 spans}; 

\node[anchor=west] (source) at (axis cs:1547,29.0){};
\node (destination) at (axis cs:1536,32){};
\draw[->](source)--(destination);

\addlegendentry{Closed-form}
\addplot[Set1-B,closedform] table[x=wavelength,y=closed] {Data/ECOC_BW_SNR_NLI_QAMxx1.txt};
\addplot[Set1-B,closedform,forget plot] table[x=wavelength,y=closed] {Data/ECOC_BW_SNR_NLI_QAMxx10.txt};

\addlegendentry{Simulation}
\addplot[Set1-A,simulation,dotted] table[x=wavelength,y=64qam_span1] {Data/ECOC_SSFM.txt};
\addplot[Set1-A,simulation,dotted] table[x=wavelength,y=64qam_span10] {Data/ECOC_SSFM.txt};


\end{axis}

\end{tikzpicture}
\caption{Nonlinear performance after 1~x~100~km and 10~x~100~km transmission for (a) Gaussian and (b) 64-QAM constellations.}
\label{fig:SNR}
\end{figure}

As the aim of this paper is to show the accuracy of the new GN-model closed-form expression accounting for arbitrary modulation formats in Raman amplified links, we only show results for $\text{SNR}_{\text{NLI},i}$.
The $\text{SNR}_{\text{NLI},i}$ is shown in Fig~\ref{fig:SNR} for (a) Gaussian constellations~\cite{arxiv} and (b) unshaped 64-QAM using the correction term in closed form shown in~Eq.\eqref{eq:NLI}. The results are shown for 1 and 10 spans. Note that, the degradation in the S-band performance is a result of the increased power levels provided by RA (see Fig~\ref{fig:profile_3D}), which increases the NLI in this band. The accuracy of the closed-form expression in Eq.~\eqref{eq:NLI} together with that in~\cite{arxiv} are validated for all the scenarios with SSFM simulations. For Gaussian constellations, the results were also validated using the integral model ~\cite{isrsgnmodel}. Over all the scenarios considered, the maximum average errors of 0.78~dB and 0.51~dB were found between the closed-form expression and the SSFM simulations, respectively for Gaussian and 64-QAM constellations, where the most accurate channels were the ones located in the S-band and, therefore, Raman amplified.   

\section{Conclusions}
The first closed-form expression supporting RA for arbitrary modulation formats is presented. The formula supports an arbitrary number of pumps and takes into account the ISRS effect, making it suitable for fast evaluation of UWB systems. The formula was applied in signal transmission with a 20~THz optical bandwidth, in which the S-band is fully amplified using backward-distributed RA. The accuracy of the formula was validated with SSFM simulations. With the proposed closed-form formula, the NLI can be calculated in microseconds using state-of-art processors.

\footnotesize{
\textbf{Funding.} This work is funded by the EPSRC Programme Grant TRANSNET (EP/R035342/1) and EWOC (EP/W015714/1), an EPSRC studentship (EP/T517793/1), the Microsoft 'Optics for the Cloud' Alliance and a UCL Faculty of Engineering Sciences Studentship.}


\printbibliography

\vspace{-4mm}

\end{document}